\documentclass[twocolumn,secnumarabic,showpacs,showkeys,
amssymb, amsmath,nobibnotes, aps, pre]{revtex4}

\usepackage{graphicx}% Include figure files
\begin{document}

%\preprint{APS/123-QED}

\title{Extremum complexity in the monodimensional ideal gas: \\
the piecewise uniform density distribution approximation}
% Force line breaks with \\

\author{Xavier Calbet}
% \altaffiliation[]{Physics Department, XYZ University.}%Lines break automatically or can be forced with \\
 \email{xcalbet@googlemail.com}
\affiliation{%
BIFI,\\
Universidad de Zaragoza,\\
E-50009 Zaragoza, Spain.\\
}%

\author{Ricardo L\'opez-Ruiz}%
% \homepage{http://www.Second.institution.edu/~Charlie.Author}
 \email{rilopez@unizar.es}
\affiliation{
DIIS and BIFI,\\
Universidad de Zaragoza,\\
E-50009 Zaragoza, Spain.\\
}%

\date{\today}% It is always \today, today,
             %  but any date may be explicitly specified

\begin{abstract}
In this work, it is suggested that the extremum complexity distribution of a high dimensional 
dynamical system can be interpreted as a piecewise uniform distribution in the phase space 
of its accessible states. When these distributions are expressed as one--particle distribution 
functions, this leads to piecewise exponential functions.
It seems plausible to use these distributions in some systems out of equilibrium, 
thus greatly simplifying their description.
In particular, here we study an isolated ideal monodimensional gas far from 
equilibrium that presents an energy distribution formed by two non--overlapping Gaussian 
distribution functions. This is demonstrated by numerical simulations. 
Also, some previous laboratory experiments with granular systems seem to display 
this kind of distributions.
\end{abstract}

\pacs{89.75.Fb, 05.45.-a, 02.50.-r, 05.70.-a}% PACS, the Physics and Astronomy
                             % Classification Scheme.
\keywords{nonequilibrium systems, ideal gas, complexity}%Use showkeys class option if keyword
                              %display desired
\maketitle

\section{Introduction}

In general, a variational formulation can be established for the principles
that govern the physical world. Thus, the technique of extremizing a particular 
physical quantity has been traditionally very useful for solving many different problems.
A notable example in the thermodynamics field
is that of the maximum entropy principle, which basically
states that a system under constraints (i.e. isolated) will evolve
by monotonically increasing its entropy with time 
and will reach equilibrium
at its maximum achievable entropy \cite{jaynes57}.
This principle is valuable in two distinct ways.
First, it unambiguously provides a way to determine
the state of equilibrium, which for the case of an isolated system
will be that of the equiprobability
among the accessible states. This property is useful
within the field of equilibrium thermodynamics or thermostatics.
And secondly, it gives a definite
direction in which the system will evolve toward equilibrium, 
which is effectively an arrow of time. This property is valuable by restricting
the evolution of systems to those of ever growing entropy.
It also tells us that the entropy is equivalent to a stretched
or compressed time axis. In summary, the latter property is useful for
thermodynamics in its broader sense, that is, for systems
out of equilibrium.

Recently \cite{calbet07},
the extremum complexity assumption has been proven valuable
for greatly restricting
the possible accessible states of an
isolated system far away from equilibrium.
It states that isolated systems
out of equilibrium can be
simplified by assuming equiprobability among some
of the total accessible states and zero probability of
occupation for the rest of them.
Equivalently, we can say that the probability
density function of the system is approximated
by a piecewise uniform distribution among the accessible states.
The spirit of this hypothesis is that in some
isolated systems local complexity can arise despite
its increase in entropy. A typical example being life
which can be maintained in an isolated system as
long as internal resources last.

In this paper, we will justify this hypothesis and
will extend this concept applied to the monodimensional
ideal gas. It will be shown that for some isolated
systems relaxing towards equilibrium, it is a good
approximation to assume that the system follows
a series of states with extremum complexity, the
{\it extremum complexity path}. The usefulness of this
idea resides in simplifying the dynamics
of the system by allowing to describe very complex
systems with just a few parameters. Advancing some of our results,
in section \ref{sec:maxcomplexapprox}, the state
of a monodimensional ideal gas far from equilibrium with $10,000$ particles
will be explained by a reduced set of only nine variables. 
We shall also see in section \ref{sec:granular} how
some experiments with granular systems \cite{kudrolli00} also 
seem to show an extremum complexity distribution.

In section \ref{sec:maxcomplex}, the equivalence between extremum complexity
states and piecewise uniform distributions will be presented.
A justification for this assumption will be
explained in section \ref{sec:just}. These concepts will be applied
to the monodimensional ideal gas in section \ref{sec:gas}. The
extremum complexity distribution and approximations in this
monodimensional ideal gas
are shown in section \ref{sec:maxcomplexgas} and
\ref{sec:maxcomplexapprox},
where
the assumption of extremum complexity will be shown
to greatly simplify the dynamics of the system.
Results of the numerical simulation of the monodimensional
ideal gas are presented in section \ref{sec:results}.
Some distributions found in experiments with granular
systems \cite{kudrolli00} also seem to be extremum complexity ones.
This is suggested in section \ref{sec:granular}.
Finally, a discussion of the results is given
in section \ref{sec:discussion}.

\section{Extremum complexity distribution is a piecewise
uniform density distribution}\label{sec:maxcomplex}

When an isolated system relaxes towards equilibrium, 
it does so by monotonically increasing its entropy. 
In this context, the entropy is equivalent to a stretched time axis.  
In some particular cases \cite{calbet01,calbet07}, the extremum complexity
hypothesis can be also assumed, which means that we have a supplementary constraint,
namely, the isolated system prefers to relax toward equilibrium by approaching 
or following the extremum complexity path \cite{calbet01}.

The so called LMC complexity $C$ given by \cite{lopez-ruiz95} is defined as

\begin{equation}
C = H \cdot D,
\end{equation}

where the disequilibrium, $D$ is defined in \cite{lopez-ruiz95} as the
distance of the system state to the microcanonical
equilibrium distribution, the equiprobability,

\begin{equation}
D = \sum_{i=1}^N ( f_i - 1/N )^2,
\end{equation}

and $H$ is the normalized entropy,

\begin{equation}
H=-(1/\ln N) \sum_{i=1}^N f_i \ln f_i,
\end{equation}

where $N$ is the number of accessible states
and $f_i$, with $i=1, 2, \dots N$, is
the probability of occurrence of particular state $i$ of the system.
Other authors have proposed different
definitions for the disequilibrium, which are claimed to exhibit a more 
appropriate behavior than the original definition for some particular applications. 
See Martin et al. \cite{martin06} for a comprehensive list of them. 
At any rate, the extremum complexity distribution happens to be identical for
almost all of these LMC-like complexities \cite{martin06}.

\begin{figure}
\includegraphics[angle=-90,width=0.9\columnwidth]{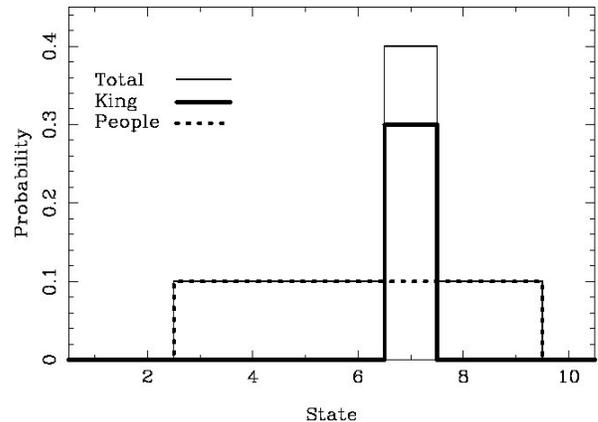}
\caption{\label{fig:maxcomp}
Microcanonical extremum complexity distribution derived
in Ref. \cite{calbet01} for $N=8$.
It is the sum of the king (thick solid line and $f_K$ with $f_k=0.3$
in Table \ref{tab:extremum}) 
and the people, or equiprobability for non-zero states, 
distribution (dotted line and $f_P$ with $f_p=0.1$ in Table \ref{tab:extremum}).}
\end{figure}

The extremum complexity distribution can be
calculated by finding the complexity extrema for
a given entropy, $H$, using Lagrange multipliers \cite{calbet01}.
Table \ref{tab:extremum} shows the resulting distribution functions.
The extremum distribution function is graphically shown
for all accessible states of the system
in Fig. \ref{fig:maxcomp}. It can be subdivided in two components, one
with the maximum probability, which will be referred
as ``king distribution", and another one with the rest of the non-zero
probabilities, which will be named ``people distribution".
An important aspect to note is that both distributions are uniform
within a certain domain of the phase space and zero everywhere
else. They are effectively piecewise uniform distributions.
In this respect, extremum complexity distributions are equivalent
to piecewise uniform ones.

\begin{table}
\caption{\label{tab:extremum} Probability values, $f_j$, 
that give a maximum
complexity, $C$,
for a given entropy, $H$. $f_k$ is the king distribution and
$f_p$ is the people distribution.}
\begin{ruledtabular}
\begin{tabular}{ccc}
Number of states with $\displaystyle f_j$
    & $\displaystyle f_j$ & Range of $\displaystyle f_j$\\
\hline
$\displaystyle 1$         & $f_k = f_{\rm max}$
         & $\displaystyle 1/N\ \ldots\ 1$\\
$\displaystyle N - 1$     & $f_p = (1 - f_{\rm max})/(N - 1)$ & 
         $\displaystyle 0\ \ldots\ 1/N$\\
\end{tabular}
\end{ruledtabular}
\end{table}

\section{Evolution of a piecewise uniform distribution}
\label{sec:just}

There are many systems where the initial probability 
distribution function is effectively
a piecewise uniform function. A clear example is the monodimensional
ideal gas with $N$ particles as shown in \cite{calbet07} and in section
\ref{sec:gas}.
In this example two very energetic particles are introduced
into a gas previously in equilibrium. The picture of the system, just
before the introduction of the extreme energetic particles,
is that of a distribution function in which the
accessible states of the system form a uniformly distributed
hypersphere in
the $(N-2)$-dimensional phase space.
Right after we introduce the high energetic particles,
the whole accessible space is blown up extremely into
a much bigger hypersphere in the final $N$-dimensional phase space.
The original $N-2$ particles that where in equilibrium will still occupy
the $(N-2)$-dimensional uniformly distributed hypersphere, that is,
a subspace of the much bigger $N$-dimensional hypersphere.
The 2 energetic particles will occupy a place in the $N$-dimensional
hypersphere with very high momentum. Because of this, globally, a very
big part of the accessible N-dimensional hypersphere will
have zero probability of being occupied, or in other words,
the global distribution could be approximated by
one or more (two for this example of the monodimensional gas) 
piecewise uniform distribution functions.

After this initial moment,
the distribution function will be modified as the system relaxes
towards equilibrium until it reaches the equiprobability within
the $N$-dimensional hypersphere.
The evolution of the system between this initial state
and the equilibrium one will depend on the particular example studied.
Still, we can sketch how the system will evolve by using
an important property of Frobenius-Perron operators \cite{mackey92}.
If we define $S$ as the nonsingular transformation governing the evolution
of a dynamical system in phase space $x$, $f$ 
as the probability density function
and $P$ as the operator to calculate the evolution of $f$, then
this property states that for every set $A$, $Pf(x)=0$ for all
elements $x \in A$ if and only if $f(x)=0$ for all
$x \in S^{-1}(A)$.
In practical terms this means that if the system is initially
with many states with zero probability of occurrence, 
then the system will evolve
by keeping many of the regions of phase space with zero probability. 
The rest of the non-zero probability states can be assigned
a constant probability as a first approximation.
In other words, the system can be approximated
as evolving
from one piecewise uniform density function to another
one, always remaining close to the extremum complexity state.

To illustrate this concept, we will look at the density
function of a system that evolves by following the tent map.
The tent map is governed by the following equations,
\begin{equation}
S(x)= \quad
\begin{cases}
2 x &, 0 \le x < \frac{1}{2}\\
2(1-x) &, \frac{1}{2} \le x \le 1
\end{cases}
\end{equation}
Its density function evolution can be easily calculated
\cite{mackey92} and is given by
\begin{equation}
Pf(x)=\frac{1}{2}\left[ f(\frac{1}{2}x)+f(1-\frac{1}{2}x)\right] 
\end{equation}

\begin{figure}
\includegraphics[angle=-90,width=0.9\columnwidth]{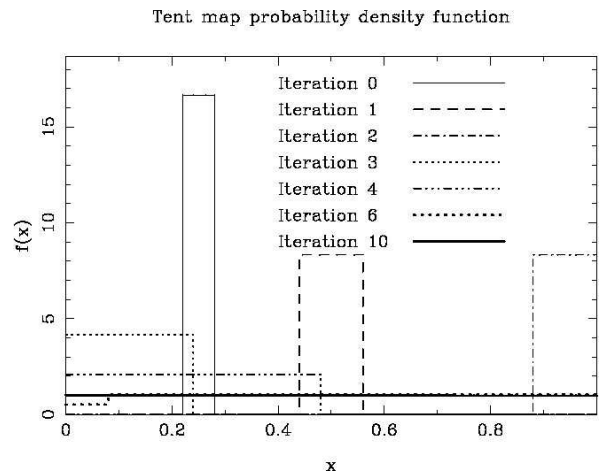}
\caption{\label{fig:tent_distri}
Time evolution of an initially piecewise uniform distribution function
for the tent map. The distribution function tends to stay, in a first
approximation, as a piecewise uniform distribution.}
\end{figure}

The stationary density function for this system is the
equiprobability.
If we now initialize the system with a piecewise uniform
density function, we will verify that the density distribution
is transformed from one piecewise uniform function to 
approximately another one
in such a way that the region with zero probability keeps
getting smaller (see Fig. \ref{fig:tent_distri}). At the final stage, 
the zero probability region
disappears and the system is in an equiprobable distribution.

Following this illustration and thanks to the above mentioned
property of the Frobenius-Perron operators,
it is to be expected that a system which has the equiprobability
as its equilibrium distribution and is initialized with
a function similar to a piecewise uniform distribution,
will evolve approximately following subsequent piecewise uniform distributions
until it reaches the equiprobability.

\section{The monodimensional ideal gas far from equilibrium}
\label{sec:gas}

In the simulations, the gas is initially,
at time $t=0$, in equilibrium. Its
one particle momentum distribution is
described by a Gaussian or Maxwell--Boltzmann function.
At this point, two new extremely energetic particles are introduced
into the gas, forcing the gas into a far
from equilibrium state.
The system is kept isolated from then on.
It eventually relaxes again toward equilibrium showing
asymptotically another Gaussian distribution.
The relaxation towards equilibrium of this isolated monodimensional
gas follows the extremum complexity path. 

The evolution of the
system can be approximated
by two functions, the people and the king distribution,
which are piecewise uniform
on the accessible states of the system \cite{calbet07}.
A double Gaussian distribution will result when we
transform these N-dimensional distributions into
one particle momentum ones (see Fig. \ref{fig:gausfit_150000}).

In more detail, and using arbitrary units from now on,
the gas consisted of $10,000$ point-like particles
colliding with each other elastically.
The particles were positioned with 
alternating masses of $1$ and $2$ on regular intervals 
on a linear space
$10,000$ units long. The system has no boundaries,
i.e., the last particle in this linear
space was allowed to collide with the first one, in a way similar
to a set of rods on a circular ring.
Two distinct masses in the system
were used because 
a monodimensional gas can thermalize only if
its constituent particles have at least two
different masses.
Initially $9998$ particles were given initial conditions
following a Gaussian distribution with mean zero
velocity and a mean energy of $1/2$, giving a
total mean energy for the system of nearly $5,000$.
These particles where then allowed to undergo $20$ million 
collisions in order for the system to reach the initial state of 
equilibrium, i.e., a Gaussian distribution.
After that, at time $t=0$, two extremely energetic particles
of mass $1$ and $2$ are introduced at two neighboring points, 
such that the total system
has zero momentum and several different final energies of 
$E=10,000$, $75,000$, $150,000$ and $1,500,000$.
The system then undergoes another $20$ million collisions 
to reach again the equilibrium Gaussian distribution.
We record the time evolution of the one-dimensional momentum distribution
in Fig. \ref{fig:gausfit_150000}, where the square of the
generalized particle momentum is given by the
variable 
\begin{equation}
 p^2_i \equiv P^2_i / m_i,
\end{equation}

with $P_i$ and $m_i$ the momentum and mass of particle $i$ respectively.
The theoretical extremum complexity or piecewise uniform distributions
(derived below) for this system is approximated by
two non-overlapping Gaussian distributions and is also fitted 
as solid lines in Fig. \ref{fig:gausfit_150000}.
Let us remark that the system stays
in this double Gaussian, the extremum complexity distribution, 
during a large part of its out of equilibrium state. 
The two clearly visible slopes of Fig. \ref{fig:gausfit_150000} 
are related with the two different widths
associated with both Gaussian distributions.
As the system approaches equilibrium, both Gaussian distributions
merge into one. More details of the numerical simulations
are deferred to section \ref{sec:results}.

\section{Extremum Complexity distribution of an isolated 
monodimensional ideal gas}
\label{sec:maxcomplexgas}

In \cite{calbet07}, we showed a different derivation of the extremum complexity 
distribution to the one shown here, both obtaining the same results.
We will make our study inside the N-dimensional phase space
constituted by the generalized momentums. We will also define an
m-dimensional sphere as one located in an m-dimensional space.

Since initially $N-2$ particles are in equilibrium, they will be located
on an $N-2$ dimensional hypersphere within the $N$ dimensional
phase space. This will be our people distribution. The
remaining 2 high energetic particles will be located on a
2-dimensional hypersphere embedded in the global N-dimensional
phase space. This will conform the king distribution.

As the system evolves in time, the particles of the people distribution
will slowly migrate to the king distribution.
Following the extremum complexity approximation, we will assume
both distributions are piecewise uniform ones, being uniform where they
are non-zero.

Each one of the particles will be assumed to be either
in the people or king distribution. The distributions
will be separated by a particular value of the generalized
momentum $p_0$. If the absolute value of the
generalized momentum is below (above)
$p_0$ then the particle will belong to the people (king) distribution.
Denoting by $n_p$, $n_k$
the number of particles and $e_p$ and $e_k$ the half--variances
of the Gaussian distributions in the
people and the king distributions, respectively , then
the particles of the people distribution, with absolute generalized momentums
below $p_0$, are distributed uniformly
over the $n_p$-dimensional sphere of radius $\sqrt{2 n_p e_p}$. 
The rest of the particles, forming the king distribution,
with absolute generalized momentums above
$p_0$, will be distributed uniformly on an $n_k$-dimensional
sphere of radius $\sqrt{2 n_k e_k}$.
The cartesian product of these hyperspheres will be embedded on the
$N$-dimensional hypersphere within the complete $N$-dimensional
phase space.

We can now calculate the one--particle momentum distribution
by integrating both distributions in $N$--dimensional
phase space into just one dimension.
This is done in Appendix \ref{sec:proof}.
Combining both one--particle momentum distributions,
the complete distribution
function for the monodimensional ideal gas far away from
equilibrium can be written as,

\begin{equation}
f(p) = \quad
\begin{cases}
K_p \; \exp ( - p^2 / 4 e_p),& |p| < p_0\\
K_k \; \exp ( - p^2 / 4 e_k),& |p| \ge p_0
\end{cases}
\end{equation}

with $K_p$ and $K_k$ normalization constants of the distribution
function, such that,

\begin{eqnarray}
\int_0^{p_0} f(p) \; {\rm d} p = n_p\\
\int_{p_0}^{\infty} f(p) \; {\rm d} p = n_k,
\end{eqnarray}

which must satisfy the conservation of particles by maintaining
$N$ constant,

\begin{equation}
N=n_p+n_k.
\end{equation}

Separating the people and the king function we can
now define,

\begin{eqnarray}
\label{eq:fpfk}
 f_p(p) \equiv K_p \; \exp ( - p^2 / 4 e_p) \nonumber \\
 f_k(p) \equiv K_k \; \exp ( - p^2 / 4 e_k).
\end{eqnarray}

We will also require the distribution function to be continuous,

\begin{equation}
f_p(p_0) = f_k(p_0).
\end{equation}

\section{Extremum complexity approximation equations in the monodimensional gas}
\label{sec:maxcomplexapprox}

We can now define the total energies, $E_p$ and $E_k$, and the mean energy per particle,
$\overline{e_p}$ and $\overline{e_k}$,
of the people and king distribution such that the total energy of the system, $E$,
is conserved,

\begin{eqnarray}
E_p = n_p \overline{e_p} \nonumber \\
E_k = n_k \overline{e_k}  \nonumber \\
E = E_p + E_k.
\end{eqnarray}

We can combine the distribution functions of Eq. (\ref{eq:fpfk}), 
the conservation principles discussed
and the concepts shown in
section \ref{sec:maxcomplexgas}
to obtain the extremum complexity approximation equations of the
monodimensional gas,

\begin{eqnarray}
n_p = \int_0^{p_0} K_p \; \exp ( - p^2 / 4 e_p) \; {\rm d} p \label{eq:np}\\
n_k = \int_{p_0}^{\infty} K_k \; \exp ( - p^2 / 4 e_k) \; {\rm d} p \label{eq:nk}\\
N = n_p + n_k \label{eq:N} \\
K_p \exp ( - p_0^2 / 4 e_p) = K_k \; \exp ( - p_0^2 / 4 e_k) \label{eq:cont}\\
E = n_p \overline{e_p} + n_k \overline{e_k} \label{eq:E}\\
\overline{e_p} = \dfrac{\int_0^{p_0} \dfrac{p^2}{2} \exp (- p^2 / 4 e_p) {\rm d} p}{\int_0^{p_0} 
\exp ( - p^2 / 4 e_p) \; {\rm d} p} \label{eq:mep}\\
\overline{e_k} = \dfrac{\int_{p_0}^{\infty} \dfrac{p^2}{2} 
\exp (- p^2 / 4 e_k) {\rm d} p}{\int_{p_0}^{\infty} \exp ( - p^2 / 4 e_k) \; {\rm d} p} \label{eq:mek},
\end{eqnarray}

where Eqs. (\ref{eq:np}) and (\ref{eq:nk}) are the expressions of the number of particles
in each distribution, Eq. (\ref{eq:N}) is the conservation of particles,
Eq. (\ref{eq:cont}) is the continuity of the distribution function,
Eq. (\ref{eq:E}) is the conservation of energy
and Eqs. (\ref{eq:mep}) and (\ref{eq:mek}) are the definitions of the mean energy per particle.

We are thus left with seven equations and nine unknowns,
$n_p$, $K_p$, $e_p$, $\overline{e_p}$, $n_k$, $K_k$, $e_k$, $\overline{e_k}$ and $p_0$,
having simplified the monodimensional ideal gas description enormously.

We need two more equations to completely describe the system.
These will be obtained from the Boltzmann integro--differential equation.
Let us first consider the two particle collisions within the monodimensional
gas, which should conserve
energy and momentum.
If the generalized momentums before the collisions are $(p,p_1)$
and the ones after
the collision are $(p',p'_1)$, and the collisions are elastic then
they must satisfy,

\begin{eqnarray}
p' = \dfrac{(m-m_1)}{(m+m_1)} p + \dfrac{2 \sqrt{m} \sqrt{m_1}}{(m+m_1)} p_1\\
p'_1 = \dfrac{2 \sqrt{m} \sqrt{m_1}}{(m+m_1)}p - \dfrac{(m-m_1)}{(m+m_1)} p_1
\end{eqnarray}

\begin{figure}
\includegraphics[angle=-90,width=0.9\columnwidth]{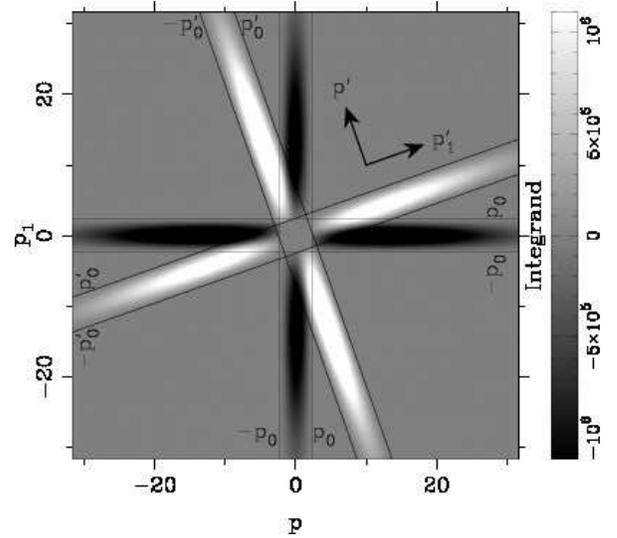}
\caption{\label{fig:integrand}Numerical values of the integrand
of the Boltzmann equation for the
monodimensional gas (Eq. (\ref{eq:boltzmann_maxcomplex})).}
\end{figure}

The net effect of a collision is a rotation and swapping of the generalized
momentums on the $(p,p_1)$ space. This is shown in Fig. \ref{fig:integrand}.
The angle of rotation $\theta$ is given by

\begin{equation}
\cos ( \theta ) = \frac{2 \sqrt{m} \sqrt{m_1} }{ m + m_1 }.
\end{equation}

Let us now obtain the Boltzmann integro--differential equation
for our particular gas.
Due to setting the gas by alternating two different masses in physical space,
for a given generalized momentum, $p$, there can correspond
particles with two different masses, $m_0$ and $m_*$.
Each one of these particles will collide with another one with
an alternative mass.
The Boltzmann equation for this particular monodimensional gas is,

\begin{align}
\label{eq:boltzmann}
\dfrac{\partial f(p)}{\partial t} =& \int \left[ \dfrac{1}{2} \left| \dfrac{p_1}{\sqrt{m_*}} - 
\dfrac{p}{\sqrt{m_0}} \right| + \dfrac{1}{2} \left| \dfrac{p_1}{\sqrt{m_0}} - 
\dfrac{p}{\sqrt{m_*}} \right| \right] \centerdot \nonumber \\
& ( f(p'_1) f(p') - f(p) f(p_1) ) \; {\rm d} p_1
\end{align}

Using the extremum complexity equations (Eqs. (\ref{eq:fpfk})) we can plot the different values
of the integrand of the Boltzmann equation to find the most significant ones.
This is shown in Fig. \ref{fig:integrand}. It can be seen that collisions in which two
particles are both in the people (or king) distribution before and after the
collision do not alter the distribution
function giving an integrand of zero.

We shall now concentrate on the evolution of the people distribution
function. As we can see from Fig. \ref{fig:integrand} the greatest contribution to the integrand
of the Boltzmann equation will come from collisions where initially
we have one particle belonging to the people distribution and the other one
to the king distribution and after the collision they both belong to the
king distribution. The original Boltzmann equation, Eq. (\ref{eq:boltzmann}),
is then simplified to,

\begin{align}
\label{eq:boltzmann_maxcomplex}
& \dfrac{\partial f_p(p)}{\partial t} \cong \int_{p_0}^{\infty} \left[ \dfrac{1}{2} 
\left| \dfrac{p_1}{\sqrt{m_*}} - \dfrac{p}{\sqrt{m_0}} \right| + \dfrac{1}{2} 
\left| \dfrac{p_1}{\sqrt{m_0}} - \dfrac{p}{\sqrt{m_*}} \right| \right] \centerdot \nonumber \\
& \left( K_k {\rm e}^{-p'^2_1/4 e_k} \; K_k {\rm e}^{-p'^{2}/4 e_k} -  K_p {\rm e}^{-p^2/4 e_p} \; 
K_k {\rm e}^{-p^2_1/4 e_k}  \right) \; {\rm d} p_1
\end{align}

Rearranging terms and using the conservation of energy, $p^{2} + p_{1}^{2} = p'^{2} + p'^{2}_{1}$, 
we are left with,

\begin{align}
&\dfrac{\partial f_p(p)}{\partial t} = \left( K_k {\rm e}^{-p^2/4 e_k} - K_p 
{\rm e}^{-p^2/4 e_p}  \right) \centerdot \nonumber \\
& \int_{p_0}^{\infty} \left[ \dfrac{1}{2} \left| \dfrac{p_1}{\sqrt{m_*}} - \dfrac{p}{\sqrt{m_0}} \right| + 
\dfrac{1}{2} \left| \dfrac{p_1}{\sqrt{m_0}} - \dfrac{p}{\sqrt{m_*}} \right| \right] \centerdot \nonumber \\
& K_k \exp(-p_1^2/4 e_k)  \; {\rm d} p_1
\end{align}

Taking into account that the momentum from the king distribution, $p_1$, is usually
much larger than the one from the people distribution, $p$, we can approximately
say that

\begin{align}
& \dfrac{1}{2} \left[ \left| \dfrac{p_1}{\sqrt{m_*}} - \dfrac{p}{\sqrt{m_0}} \right| + \dfrac{1}{2} \left| \dfrac{p_1}{\sqrt{m_0}} - \dfrac{p}{\sqrt{m_*}} \right| \right] \approx \nonumber \\
& \dfrac{1}{2} \left[ \left| \dfrac{p_1}{\sqrt{m_*}} \right| + \dfrac{1}{2} \left| \dfrac{p_1}{\sqrt{m_0}} \right| \right].
\end{align}

The probabilities of the king distribution are much lower than the ones from the
people distribution, which allows us to simplify,

\begin{align}
& \left( K_k {\rm e}^{-p^2/4 e_k} - K_p {\rm e}^{-p^2/4 e_p}  \right) \approx \nonumber \\
& - K_p \exp (-p^2/4 e_p).
\end{align}

Combining these two approximations gives us the final evolution equation
for the people distribution,

\begin{align}
\label{eq:fpevol}
\dfrac{\partial f_p(p)}{\partial t} \approx - f_p(p) \; \mu_{k},
\end{align}

where,

\begin{equation}
\mu_k = \int_{p_0}^{\infty} \left[ \dfrac{1}{2} \left| \dfrac{p_1}{\sqrt{m_*}} \right| + \dfrac{1}{2} \left| \dfrac{p_1}{\sqrt{m_0}}  \right| \right]
K_k {\rm e}^{-p_1^2/4 e_k}  \; {\rm d} p_1.
\end{equation}

Note that $\mu_k$ is similar to an average of the absolute value of the velocity times the number of particles
in the king distribution.

To transform Eq. (\ref{eq:fpevol}) into the variables used in the extremum complexity approximation,
we can integrate it for values of $p$ between $0$ and $p_0$ to obtain,

\begin{equation}
\label{eq:npevol}
\dfrac{\partial n_p}{\partial t} \simeq - n_p \; \mu_k.
\end{equation}

We can obtain an expression for the people energy by multiplying
both sides of Eq. (\ref{eq:fpevol}) by $p^2/2$ and integrating again within the limits
of the people distribution,

\begin{equation}
\label{eq:Epevol}
\dfrac{\partial E_p}{\partial t} \simeq - E_p \; \mu_k.
\end{equation}

Combining Eq. (\ref{eq:npevol}) and (\ref{eq:Epevol}) 
we can obtain the relationship of the mean
energy per particle of the people distribution,

\begin{equation}
\label{eq:mepevol}
\dfrac{\partial \overline{e_p}}{\partial t} \approx 0.
\end{equation}

Adding Eqs. (\ref{eq:npevol}) and (\ref{eq:mepevol}) to the previous set of equations
(Eqs. (\ref{eq:np}) to (\ref{eq:mek})) completes the set by having a total
of nine equations for nine unknowns.

\section{Results of the numerical simulations}
\label{sec:results}

\begin{figure}
\includegraphics[angle=-90,width=0.9\columnwidth]{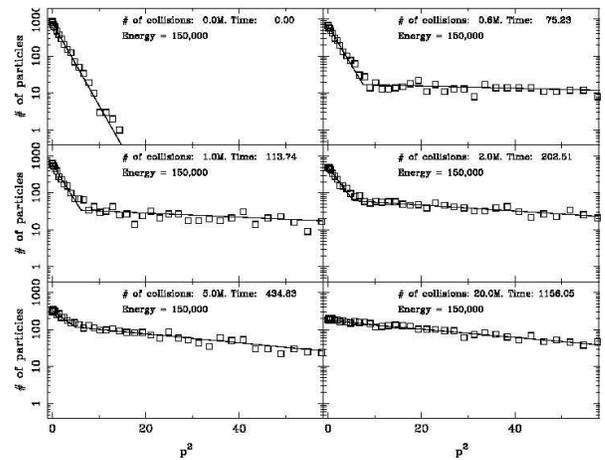}
\caption{\label{fig:gausfit_150000}Numerical simulations of an isolated
monodimensional ideal gas
where two highly energetic particles are introduced into a thermalised state
with an initial energy of about $5,000$, giving a total constant energy
of $150,000$. The logarithm of the generalized momentum histogram (squares) 
is shown as a function
of $p^2$ for different times. Two non--overlapping Gaussian functions
are fitted to the histogram and are shown as a solid line.}
\end{figure}

Numerical simulations have been carried out for $10,000$ particles
located in a space $10,000$ units long
with an energy of about $5,000$ before the introduction of the
extremely energetic particles.
This energy was distributed following a Gaussian distribution
in generalized momentum space.
To make sure they were really in an equilibrium state before the experiment started
they were further
collided 20 million times.
After this, two high energetic particles are introduced in the system
giving a total momentum of $0$.
Results for
total constant
energies of $10,000$, $75,000$, $150,000$ and $1,500,000$ after the
introduction of the energetic particles are shown
in Figs. \ref{fig:gausfit_10000}, \ref{fig:gausfit_75000},
\ref{fig:gausfit_150000} and \ref{fig:gausfit_1500000}.

For each one of these experiments, the generalized momentum histogram
has been fitted to two non-overlapping Gaussians.
There will be cases where this fit will not be easy to achieve,
namely when predominantly only one Gaussian distribution is present.
This happens at the beginning, where we mainly have the initial
Maxwell-Boltzmann distribution, and at the end of the experiment,
where we mainly have the final equilibrium Gaussian distribution.
The case with an energy of $10,000$ is a particularly difficult one to separate
at all times due to the small relative difference between the two Gaussians.
This difficulty will manifest itself as less well
determined parameters exhibiting a higher ``noise''.
Because of this, only figures of the fitted parameters with a total energy
of $1,500,000$ are shown. Results for the other energies are similar
albeit showing a higher noise.
Note that this noise could be lowered if the algorithm to fit the two
Gaussians to the experimental points were improved.

Other numerical simulations (not shown here) have been performed where the initial pair
of high energetic particles where introduced in different places
in the monodimensional gas. They all exhibit the same behavior as the ones
shown here.

\begin{figure}
\includegraphics[angle=-90,width=0.9\columnwidth]{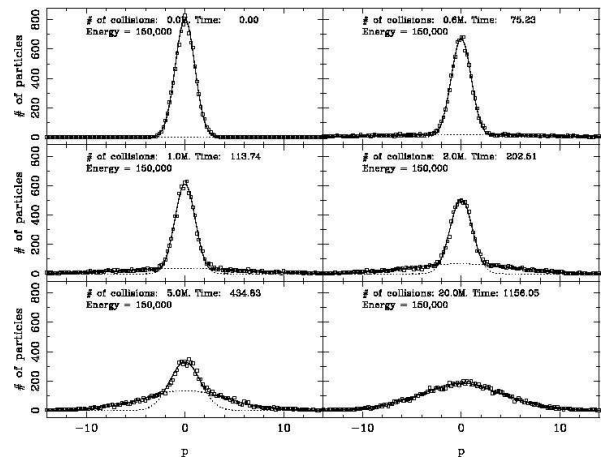}
\caption{\label{fig:gausfit_direct_150000}Numerical simulations of an isolated
monodimensional ideal gas
where two highly energetic particles are introduced into a thermalized state
with an initial energy of about $5,000$, giving a total constant energy
of $150,000$. The direct generalized momentum histogram (squares)
is shown for different times. Two non--overlapping Gaussian functions
are fitted to the histogram and are shown as a solid line.}

\end{figure}

\begin{figure}
\includegraphics[angle=-90,width=0.9\columnwidth]{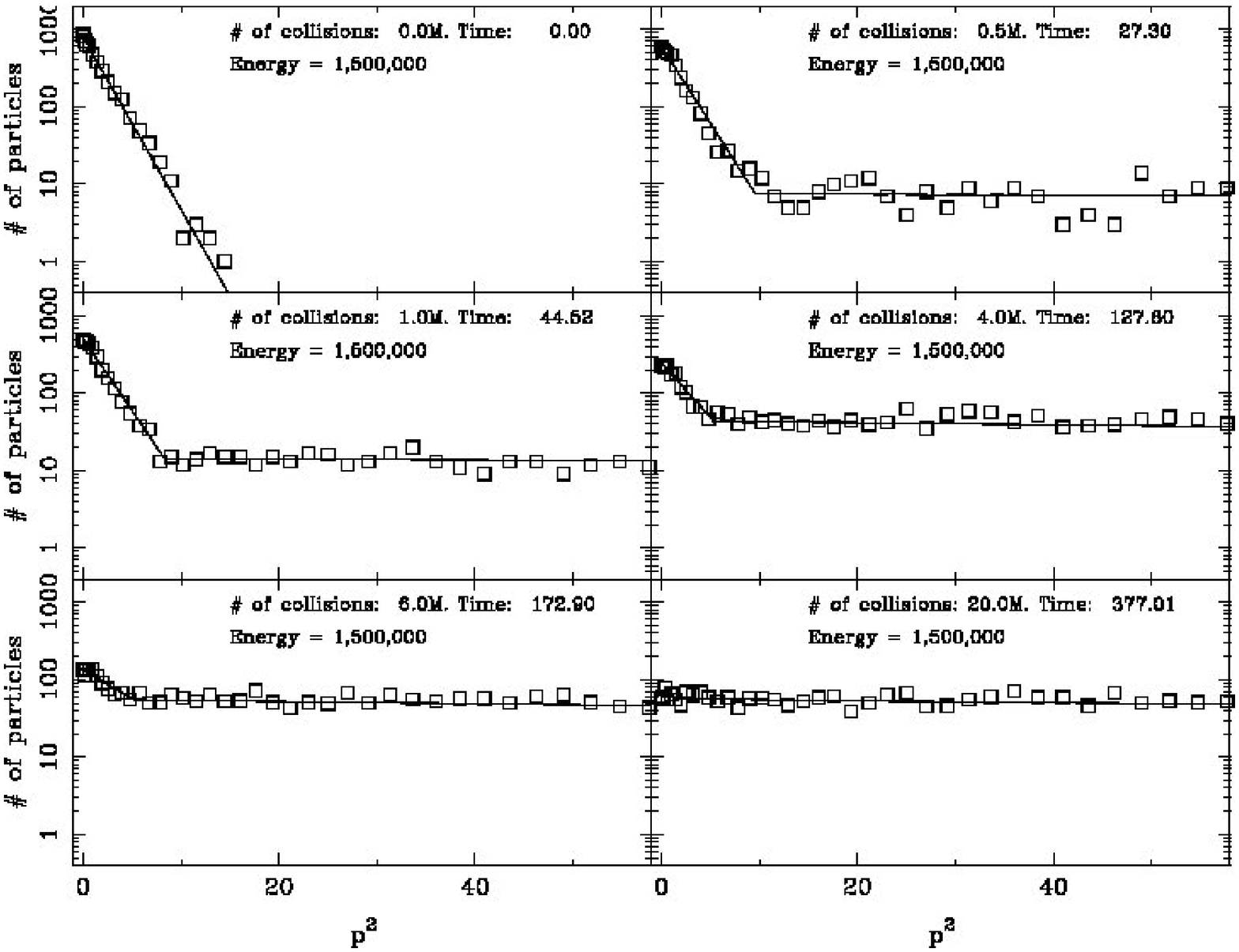}
\caption{\label{fig:gausfit_1500000}
Same as Fig. \ref{fig:gausfit_150000} but with a total constant energy of
$1,500,000$.}
\end{figure}

\begin{figure}
\includegraphics[angle=-90,width=0.9\columnwidth]{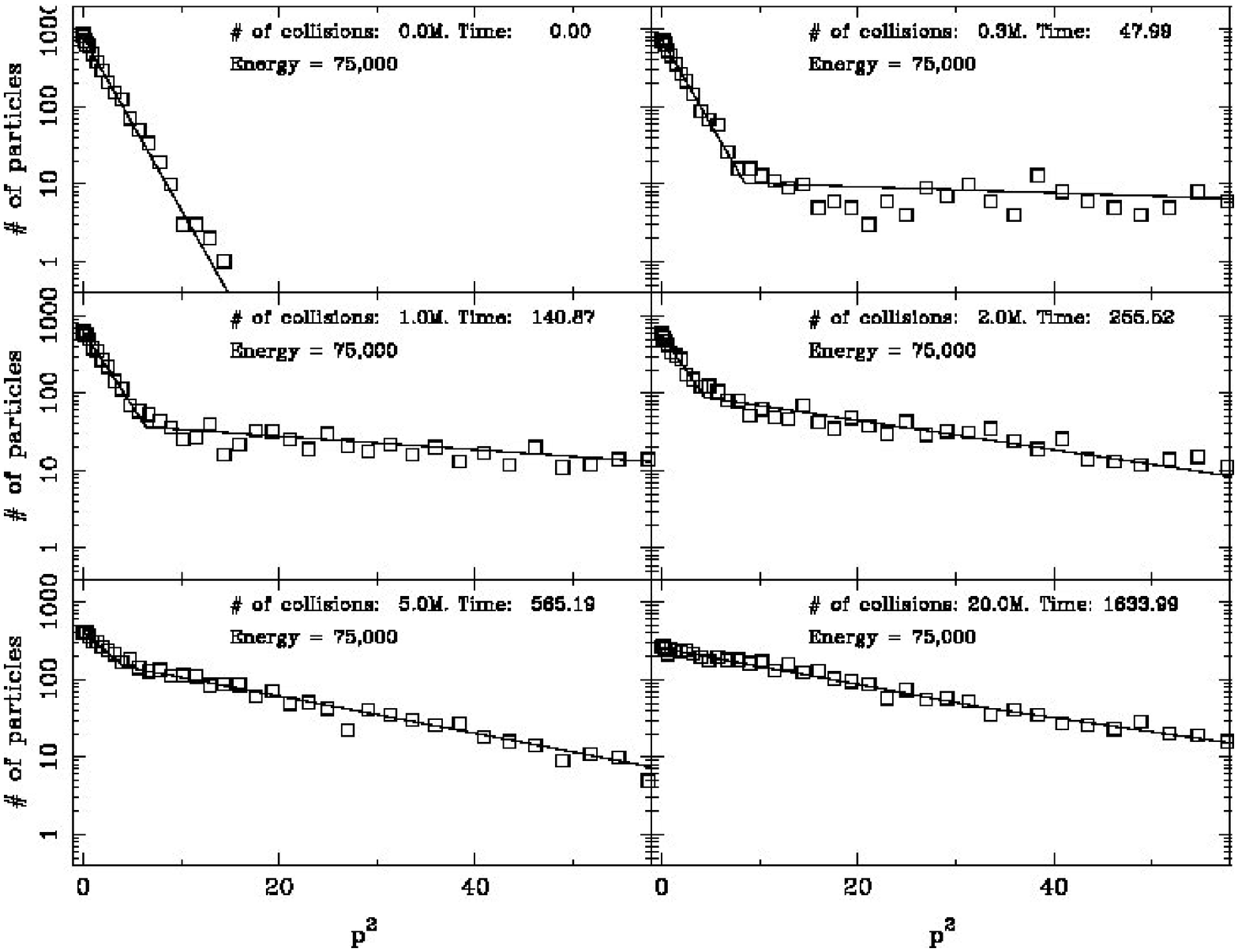}
\caption{\label{fig:gausfit_75000}Same as Fig. \ref{fig:gausfit_150000} but with a total constant energy of
$75,000$.}
\end{figure}

\begin{figure}
\includegraphics[angle=-90,width=0.9\columnwidth]{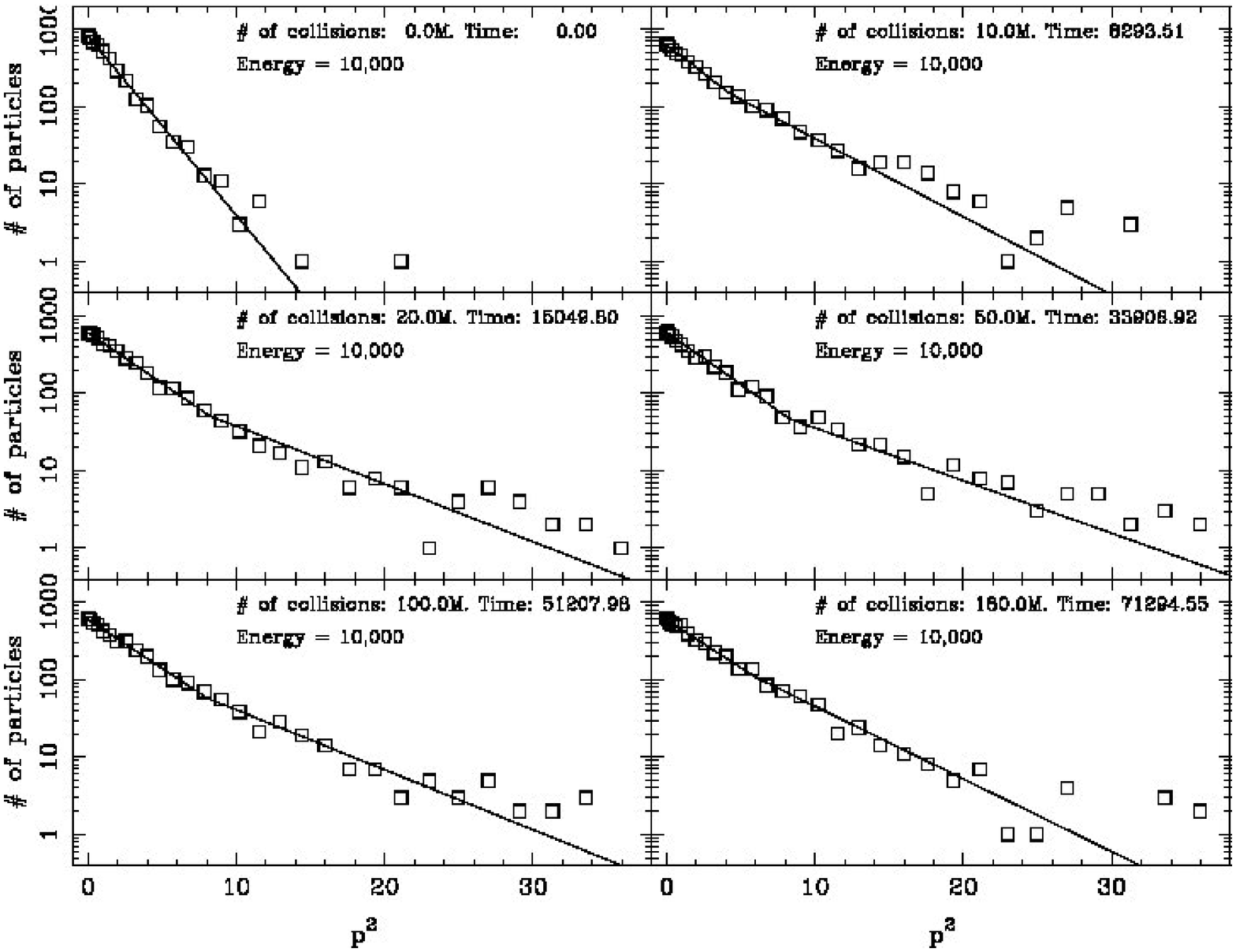}
\caption{\label{fig:gausfit_10000}Same as Fig. \ref{fig:gausfit_150000} but with a total constant energy of
$10,000$.}
\end{figure}

In Fig. \ref{fig:gausfit_150000} the results for the numerical simulations 
with a total constant energy of $150,000$ after the introduction of the
very energetic particles are shown. The axes of the graph are the generalized momentum squared
and the logarithm of the generalized
momentum.
The two non-overlapping Gaussian distributions are clearly shown.
Fig. \ref{fig:gausfit_direct_150000} shows the same histograms but in
direct scales in both axes.
Again both non-overlapping Gaussian distributions are clearly seen.

In Fig. \ref{fig:gausfit_1500000}, \ref{fig:gausfit_75000} and \ref{fig:gausfit_10000}
results with the exactly the same initial conditions (initial energy of about $5,000$) but with different 
energies
given to the extremely energetic particles providing a total constant energy of
$1,500,000$ , $75,000$ and $10,000$ respectively.
In all cases the system behaves following a double Gaussian or maximum complexity distribution.
The case with a total energy of $10,000$ is worth mentioning. This system
takes a much longer time to relax to equilibrium than the others.
It is also very difficult to separate both Gaussian distributions exhibiting
a very high noise in the derived parameters. It may
well be that in fact the system is not following the two Gaussian distribution,
but it is difficult to tell (Fig. \ref{fig:gausfit_10000}).

\begin{figure}\ref{fig:gausfit_10000}
\includegraphics[angle=-90,width=0.9\columnwidth]{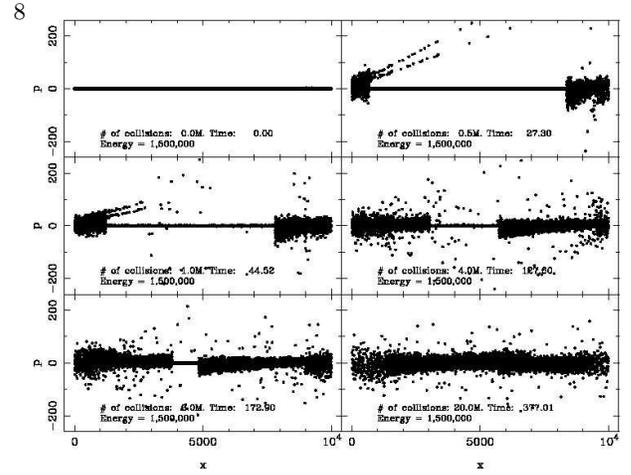}
\caption{\label{fig:phase_1500000}Monodimensional ideal gas particles in phase space.
The $x$ axis is the regular spatial coordinate and the vertical axis is the generalized momentum one.}
\end{figure}

The evolution of the system in phase space for
a total energy of 1,500,000 is shown in Fig. \ref{fig:phase_1500000}.
We can see how the gas forms clusters of higher and lower velocity particles.
Some high energetic particles are located within the low velocity clusters.

\begin{figure}
\includegraphics[angle=-90,width=0.9\columnwidth]{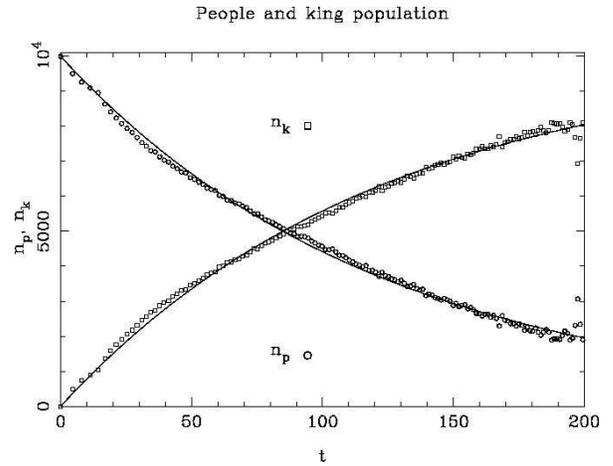}
\caption{\label{fig:n_1500000}Number of particles in the people (circles) and king (squares) distributions
as a function of time for the case with a total constant energy of $1,500,000$.
The solid line for $n_p$ is an exponential fit to the simulations,
it can be derived from Eq. (\ref{eq:np}).}
\end{figure}

\begin{figure}
\includegraphics[angle=-90,width=0.9\columnwidth]{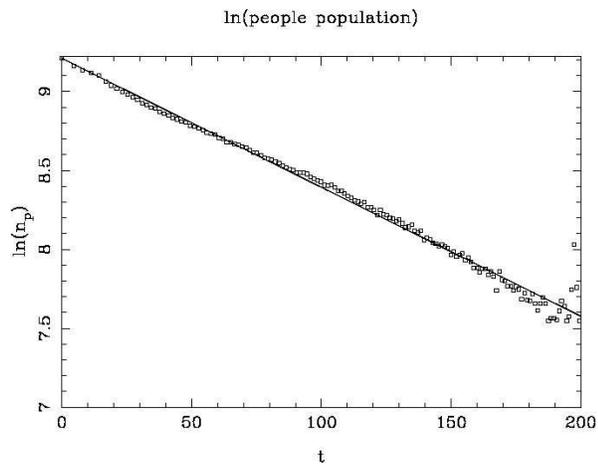}
\caption{\label{fig:log_n_1500000}Logarithm of the number of particles 
in the people distribution as a function of time for the case with a total constant energy of $1,500,000$.}
\end{figure}

The number of particles in the people and king distribution is shown in
Fig. \ref{fig:n_1500000} (energy of $1,500,000$). The difficulty in separating both Gaussian
distributions is clearly seen as ``noise'' in the figure close to
$t=0$. At higher times (right part of the graph) the noise slowly increases
until it is so high that the results at not valuable any more.
Both population numbers show an exponential evolution in time.
The solid line for $n_p$ is an exponential fit to the simulations and
it can be derived from Eq. (\ref{eq:np}).
This is better seen in Fig. \ref{fig:log_n_1500000} where the logarithm
of the people distribution population is shown. The deviation of the
exponential behavior at the right side of the graph could
be caused by the above mentioned ``noise''. Again,
the solid line is a numerical fit to the simulations.

\begin{figure}
\includegraphics[angle=-90,width=0.9\columnwidth]{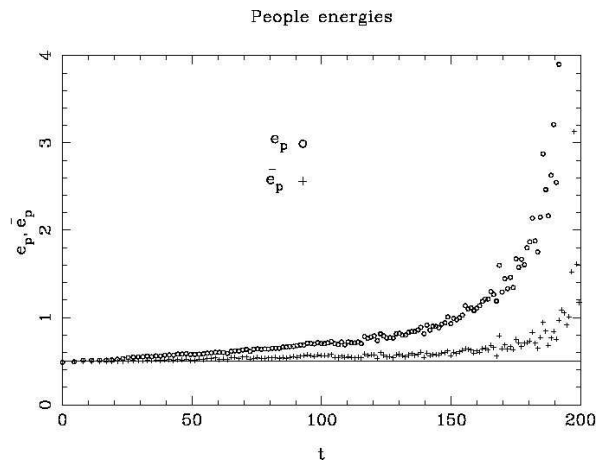}
\caption{\label{fig:e_p_1500000}Two times sigma of the Gaussian people distribution, $e_p$ (see Eq. (\ref{eq:fpfk})), 
(circles) and mean people distribution energy per particle, $\overline{e_p}$, (pluses) as a function
of time for the case with a total constant energy of $1,500,000$.
The solid line is the mean energy
per particle for the people distribution, $\overline{e_p}$, which
remains approximately constant
as expected (Eq. (\ref{eq:mepevol}).}
\end{figure}

\begin{figure}
\includegraphics[angle=-90,width=0.9\columnwidth]{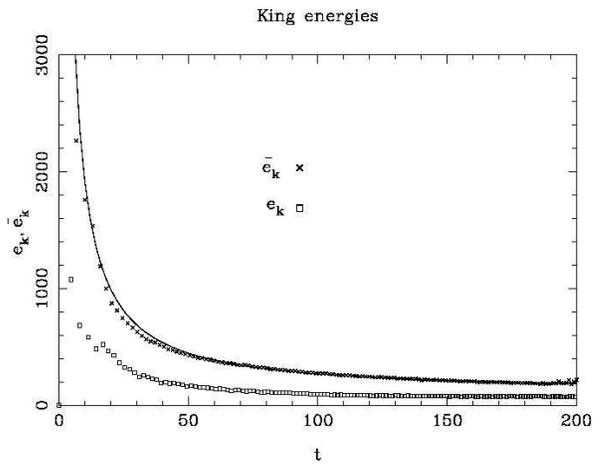}
\caption{\label{fig:e_k_1500000}Two times sigma of the Gaussian king distribution, 
$e_k$ (see Eq. (\ref{eq:fpfk})), (squares) and
mean king distribution energy per particle, $\overline{e_k}$, (crosses) as a function
of time for the case with a total constant energy of $1,500,000$.
The solid line is the mean energy per particle for the king distribution, $\overline{e_k}$, 
which varies following Eq. (\ref{eq:mek}).}
\end{figure}

The modelled parameters $e_p$ and $e_k$
parameters for the people and king distribution
using Eqs. (\ref{eq:mepevol}) and (\ref{eq:E}) as a function of time are shown in
Figs. \ref{fig:e_p_1500000} and \ref{fig:e_k_1500000} respectively.
These parameters are equivalent to the sigma of the Gaussian distributions.
The mean energy
per particle for the people distribution, $\overline{e_p}$, which
is represented as a solid line, remains approximately constant
as expected (Eq. (\ref{eq:mepevol})).

\begin{figure}
\includegraphics[angle=-90,width=0.9\columnwidth]{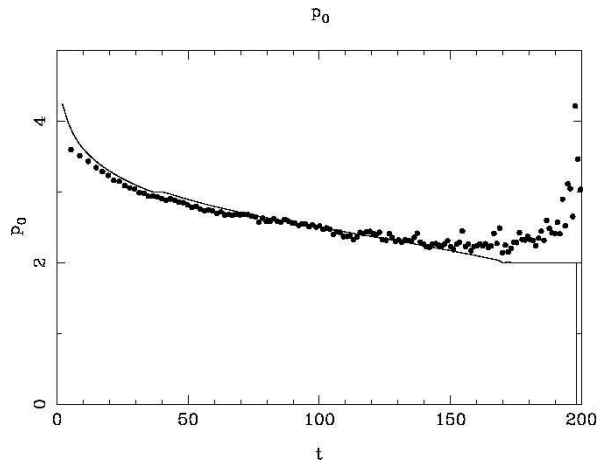}
\caption{\label{fig:p0_1500000}Momentum value, $p_0$, separating the
people and king distribution as a function of time for the case with a total constant 
energy of $1,500,000$. The solid line is the theoretical value obtained with
Eqs. (\ref{eq:np}) to (\ref{eq:mek}) and Eqs. (\ref{eq:npevol}) and (\ref{eq:mepevol}).
}
\end{figure}

In Fig. \ref{fig:p0_1500000} the momentum value, $p_0$, separating the
people and king distribution as a function of time is shown.
The solid line is the theoretical value obtained with
Eqs. (\ref{eq:np}) to (\ref{eq:mek}) and Eqs. (\ref{eq:npevol}) and (\ref{eq:mepevol}).

\begin{figure}
\includegraphics[angle=-90,width=0.9\columnwidth]{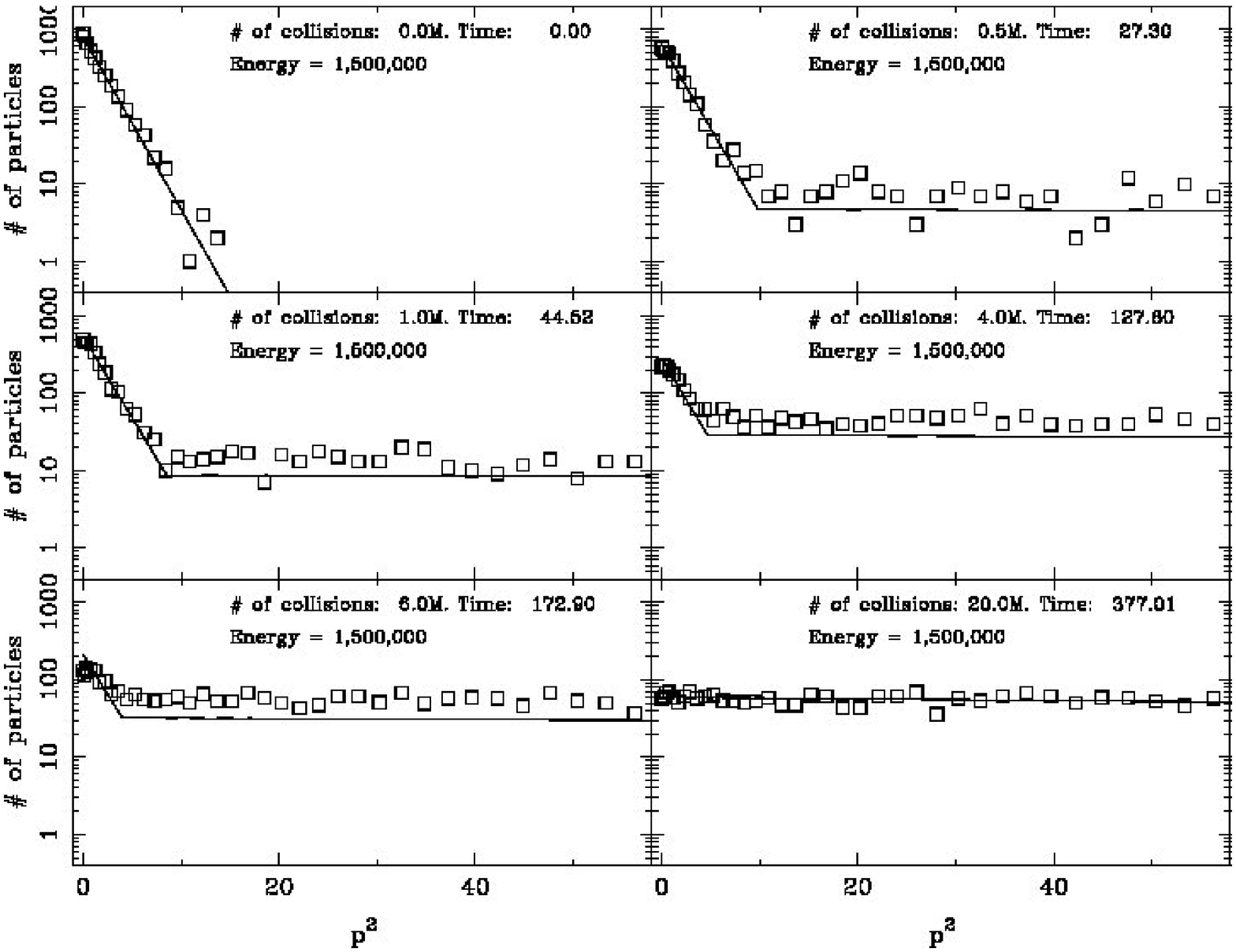}
\caption{\label{fig:gausfit_teo_1500000}
Same as Fig. \ref{fig:gausfit_150000} but with a total constant energy of
$1,500,000$. The solid lines in this cases have been obtained with the
experimental fit of $n_p$ (Fig. \ref{fig:log_n_1500000}) and 
Eqs. (\ref{eq:np}) to (\ref{eq:mek}) and Eqs. (\ref{eq:npevol}) and (\ref{eq:mepevol}).
}
\end{figure}

Finally, in Fig. \ref{fig:gausfit_teo_1500000}  the histograms for different times
are shown for a total constant energy of
$1,500,000$. In this case, the solid lines have been obtained with the
experimental fit of $n_p$ (Fig. \ref{fig:log_n_1500000}) and 
Eqs. (\ref{eq:np}) to (\ref{eq:mek}) and Eqs. (\ref{eq:npevol}) and (\ref{eq:mepevol}).
The fit could be improved (not shown in this paper) if we take second
order terms in Eqs. (\ref{eq:npevol}) and (\ref{eq:mepevol}).


\section{Granular systems in presence of gravity seen 
as extremum complexity gases}
\label{sec:granular}

\begin{figure}
\includegraphics[angle=-90,width=0.9\columnwidth]{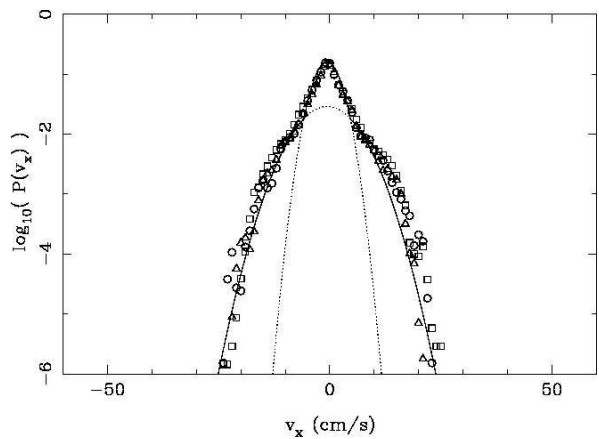}
\caption{\label{fig:granular}Non--overlapping Gaussian or
extremum complexity distribution function (solid lines) which would match
the experimental results of a granular system in the presence of gravity (
symbols as in Fig. 2 of Kudrolli and Henry \cite{kudrolli00}).}
\end{figure}

Granular system in presence of gravity experiments
consist of metal spheres allowed to move on a flat
surface which is slightly tilted with respect to the horizontal
plane and are thus subject to the force of gravity.
The particles are forced to move by an oscillation bottom wall where they
tend to fall once they dissipate enough energy by collisions.
In these systems, it has been observed that the horizontal component velocity distribution
of the particles is not Gaussian as would be expected in a system in equilibrium
\cite{kudrolli00} \cite{rouyer00} \cite{brey03}.
Although the theoretical details are not analyzed in this paper, the
system is actually in a steady state mode and could be considered
as being out of equilibrium.
These experiments and simulations show diverse velocity distributions.
Rouyer and Menon \cite{rouyer00} show a distribution with an
exponential behavior but an exponent different than two ($1.5$) as would
correspond to a Gaussian distribution.
Kudrolli and Henry \cite{kudrolli00}
show that the central portion of the velocity
distribution is Gaussian for their experiments.
Brey and Ruiz-Montero \cite{brey03} numerical simulations show neither
of them, but rather something in between depending on some properties
of the system.
In fact, some of these experiments are actually showing a double Gaussian or
extremum complexity distribution
function.
This is the case in Fig. 2 of Kudrolli and Henry \cite{kudrolli00}, where
a clear double Gaussian distribution function is shown.
To make this more evident, we present in Fig. \ref{fig:granular}
the experimental points of \cite{kudrolli00} and two fitted non-overlapping Gaussians.

\section{Discussion}
\label{sec:discussion}

The extremum complexity approximation provides a simple, yet
powerful way to simplify the dynamics of some systems out of equilibrium.
This approximation relies on a
property of the Frobenius--Perron operator
which basically keeps regions of accessible phase space
with zero probability if and only if the regions they originate from
have also zero probability.
This is the same as saying that systems out of equilibrium tend to prefer
piecewise uniform distributions in the accessible phase space.
The final consequence is that one particle distribution functions
will behave as piecewise exponential functions for some systems,
which makes the approximation relatively easy to apply to any system.
We are only left with the task of delimiting the phase space regions
for each one of the exponential distributions, the people and the king
one.
In this paper we have shown how a monodimensional gas consisting of
$10,000$ particles can be described by just nine parameters providing
a significant simplification.

One caveat of the extremum complexity approximation is that it seems to work
for some particular systems. Then, one of the consequences of this
is that it is not completely clear to which systems this method can be applied.
Another drawback is that it is not always clear how to separate the
different piecewise uniform distribution functions.
In any case, the ultimate justification of approximations in statistical mechanics
has always been that the results are useful in the real world.
This can certainly be applied to the example shown here,
the monodimensional ideal gas is enormously simplified and
accurately described by using
the extremum complexity approximation.
Also some laboratory experiments with granular 
systems out of equilibrium, as that shown in Fig. \ref{fig:granular},
display a double Gaussian distribution.

\acknowledgments

We wish to acknowledge the support of Arshad Kudrolli for providing
us with his granular system data.

The plots in this paper have
been generated with PDL (Perl Data Language, http://pdl.perl.org)

\appendix
\section{Derivation of the non-equilibrium distributions}
\label{sec:proof}

Strictly speaking,
the derivation of both non--equilibrium one--particle distributions (people and king)
should be done using the accessible states of the system, which constitutes
a surface in the N--dimensional phase space for an isolated system.
In practice, it is far easier to make this calculation using the volume
instead of the surface of the body that delimits the accessible states.
We can readily switch from volumes to surfaces using the well known
property that the volume of a sphere is proportional to its surface.

The proof will follow the technique shown in \cite{lopez-ruiz07}.
If $x_i$ are the generalized coordinates or momentums of the
Hamiltonian for particle $i$, $b$ is a parameter that defines the energy expression
for the particle and the Hamiltonian for non-interacting particles
is

\begin{equation}
 H = x_{1}^{b} + x_{2}^{b} + \dots + x_{N}^{b},
\end{equation}

then the accessible states of the system are defined by,

\begin{equation}
H \leq E,
\end{equation}

where $E$ is the total energy the system can achieve.
We will define the phase space volume inside the $E$ hypersurface
by

\begin{equation}
 V_{N} ( \rho ),
\end{equation}

being $\rho$ the maximum value the generalized coordinate
$x$ can achieve.

To this we need to add the limitations in phase
space due to the non-equilibrium condition, piecewise
uniform distribution or extremum complexity,

\begin{equation}
\label{eq:xineq}
 x_{min} \leq x_{i} \leq x_{MAX}.
\end{equation}

The relationship between the N--dimensional volume in phase
space and the volume in (N-1)-dimensional space is,

\begin{equation}
 V_{N} ( E^{\frac{1}{b}} ) = \int_{x_{min}}^{x_{MAX}} V_{N-1} ( (E-x^{b})^{\frac{1}{b}} ) dx.
\end{equation}

The one particle distribution function, $f(x)$, will be proportional to the accessible
states of the system,

\begin{equation}
\label{eq:fx}
 f(x) = C V_{N-1} ( (E-x^{b})^{\frac{1}{b}} ),
\end{equation}

where $C$ is a constant. Taking into account that $f(x)$ should be normalized,

\begin{equation}
\label{eq:fnorm}
 \int_{x_{min}}^{x_{MAX}} f(x) dx = 1,
\end{equation}

and combining Eq. (\ref{eq:fx}) and Eq. (\ref{eq:fnorm}) 
we get the following expression for $f(x)$,

\begin{equation}
\label{eq:fx2}
 f(x)=\frac{V_{N-1} ( (E-x^{b})^{\frac{1}{b}} )}{V_{N} (E^{\frac{1}{b}})}.
\end{equation}

Since the volume of an N-dimensional body as the one considered here is given
by,

\begin{equation}
 V_{N} (\rho) = g(N) \rho^{N},
\end{equation}

we can substitute this in Eq. (\ref{eq:fx2}) and arranging terms we finally get,

\begin{equation}
\label{eq:fx3}
 f(x) = \frac{g(N-1)}{g(N)} \left( 1 - \frac{x^{b}}{E} \right) ^{\frac{N-1}{b}}.
\end{equation}

We know that the total maximum energy of the system, $E$, is proportional to
the number of particles, $N$,

\begin{equation}
 E = N \epsilon.
\end{equation}

Note that because of the inequality of Eq. (\ref{eq:xineq}), $\epsilon$
will not be the mean energy per particle of the system.
Introducing this last expression in Eq. (\ref{eq:fx3}) we obtain,

 \begin{equation}
 f(x) = \frac{g(N-1)}{g(N)} \left( 1 - \frac{x^{b}}{N \epsilon} \right) ^{\frac{N-1}{b}},
\end{equation}

and finally taking the limit for $N \rightarrow \infty$ we obtain the final expression,

\begin{equation}
 f(x) = \quad
\begin{cases}
K {\rm e}^{-\frac{x^{b}}{\epsilon b}} &, \; x_{min} \leq x \leq x_{MAX}\\
0  &, \; x \leq x_{min} \;{\rm or}\; x \geq x_{MAX}
\end{cases}
\end{equation}

\bibliography{esferas16}

\begin{thebibliography}{10}
\expandafter\ifx\csname natexlab\endcsname\relax\def\natexlab#1{#1}\fi
\expandafter\ifx\csname bibnamefont\endcsname\relax
  \def\bibnamefont#1{#1}\fi
\expandafter\ifx\csname bibfnamefont\endcsname\relax
  \def\bibfnamefont#1{#1}\fi
\expandafter\ifx\csname citenamefont\endcsname\relax
  \def\citenamefont#1{#1}\fi
\expandafter\ifx\csname url\endcsname\relax
  \def\url#1{\texttt{#1}}\fi
\expandafter\ifx\csname urlprefix\endcsname\relax\def\urlprefix{URL }\fi
\providecommand{\bibinfo}[2]{#2}
\providecommand{\eprint}[2][]{\url{#2}}

\bibitem[{\citenamefont{Jaynes}(1957)}]{jaynes57}
\bibinfo{author}{\bibfnamefont{J.~T.} \bibnamefont{Jaynes}},
  \bibinfo{journal}{Phys. Rev. E} \textbf{\bibinfo{volume}{106}},
  \bibinfo{pages}{620} (\bibinfo{year}{1957}).

\bibitem[{\citenamefont{Calbet and Lopez-Ruiz}(2007)}]{calbet07}
\bibinfo{author}{\bibfnamefont{X.}~\bibnamefont{Calbet}} \bibnamefont{and}
  \bibinfo{author}{\bibfnamefont{R.}~\bibnamefont{Lopez-Ruiz}},
  \bibinfo{journal}{Physica A} \textbf{\bibinfo{volume}{382}},
  \bibinfo{pages}{523} (\bibinfo{year}{2007}).

\bibitem[{\citenamefont{Kudrolli and Henry}(2000)}]{kudrolli00}
\bibinfo{author}{\bibfnamefont{A.}~\bibnamefont{Kudrolli}} \bibnamefont{and}
  \bibinfo{author}{\bibfnamefont{J.}~\bibnamefont{Henry}},
  \bibinfo{journal}{Phys. Rev. E} \textbf{\bibinfo{volume}{62}},
  \bibinfo{pages}{R1489} (\bibinfo{year}{2000}).

\bibitem[{\citenamefont{Calbet and Lopez-Ruiz}(2001)}]{calbet01}
\bibinfo{author}{\bibfnamefont{X.}~\bibnamefont{Calbet}} \bibnamefont{and}
  \bibinfo{author}{\bibfnamefont{R.}~\bibnamefont{Lopez-Ruiz}},
  \bibinfo{journal}{Phys. Rev. E} \textbf{\bibinfo{volume}{63}},
  \bibinfo{pages}{066116(9)} (\bibinfo{year}{2001}).

\bibitem[{\citenamefont{Lopez-Ruiz et~al.}(1995)\citenamefont{Lopez-Ruiz,
  Mancini, and Calbet}}]{lopez-ruiz95}
\bibinfo{author}{\bibfnamefont{R.}~\bibnamefont{Lopez-Ruiz}},
  \bibinfo{author}{\bibfnamefont{H.}~\bibnamefont{Mancini}}, \bibnamefont{and}
  \bibinfo{author}{\bibfnamefont{X.}~\bibnamefont{Calbet}},
  \bibinfo{journal}{Physics Letters A} \textbf{\bibinfo{volume}{209}},
  \bibinfo{pages}{321} (\bibinfo{year}{1995}).

\bibitem[{\citenamefont{Martin et~al.}(2006)\citenamefont{Martin, Plastino, and
  Rosso}}]{martin06}
\bibinfo{author}{\bibfnamefont{M.}~\bibnamefont{Martin}},
  \bibinfo{author}{\bibfnamefont{A.}~\bibnamefont{Plastino}}, \bibnamefont{and}
  \bibinfo{author}{\bibfnamefont{O.}~\bibnamefont{Rosso}},
  \bibinfo{journal}{Physica A} \textbf{\bibinfo{volume}{369}},
  \bibinfo{pages}{439} (\bibinfo{year}{2006}).

\bibitem[{\citenamefont{Mackey}(1992)}]{mackey92}
\bibinfo{author}{\bibfnamefont{M.~C.} \bibnamefont{Mackey}},
  \bibinfo{journal}{Springer-Verlag, Berlin} pp. \bibinfo{pages}{24--25,40--41}
  (\bibinfo{year}{1992}).

\bibitem[{\citenamefont{Rouyer and Menon}(2000)}]{rouyer00}
\bibinfo{author}{\bibfnamefont{F.}~\bibnamefont{Rouyer}} \bibnamefont{and}
  \bibinfo{author}{\bibfnamefont{N.}~\bibnamefont{Menon}},
  \bibinfo{journal}{Phys. Rev. Lett.} \textbf{\bibinfo{volume}{85}},
  \bibinfo{pages}{3676} (\bibinfo{year}{2000}).

\bibitem[{\citenamefont{Brey and Ruiz-Montero}(2003)}]{brey03}
\bibinfo{author}{\bibfnamefont{J.~J.} \bibnamefont{Brey}} \bibnamefont{and}
  \bibinfo{author}{\bibfnamefont{M.~J.} \bibnamefont{Ruiz-Montero}},
  \bibinfo{journal}{Phys. Rev. E} \textbf{\bibinfo{volume}{67}},
  \bibinfo{pages}{021307} (\bibinfo{year}{2003}).

\bibitem[{\citenamefont{Lopez-Ruiz et~al.}(2007)\citenamefont{Lopez-Ruiz,
  Sanudo, and Calbet}}]{lopez-ruiz07}
\bibinfo{author}{\bibfnamefont{R.}~\bibnamefont{Lopez-Ruiz}},
  \bibinfo{author}{\bibfnamefont{J.}~\bibnamefont{Sanudo}}, \bibnamefont{and}
  \bibinfo{author}{\bibfnamefont{X.}~\bibnamefont{Calbet}},
  \bibinfo{journal}{arXiv:0708.3761v2 [nlin.CD]}  (\bibinfo{year}{2007}).

\end{thebibliography}

\end{document}